\documentclass[aps,twocolumn,superscriptaddress,showpacs]{revtex4}

\usepackage[dvips]{graphicx}

\usepackage[]{caption}

\newcommand{\kb}{k_{\rm B}}
\newcommand{\lb}{\ell_{\rm B}}
\newcommand{\lbt}{\widetilde{\ell}_{\rm B}}
\newcommand{\til}[1]{\widetilde{#1}}

\begin{document}

\title{Polyelectrolyte-colloid complexes: polarizability and effective
  interaction}

\author{J.~Dzubiella}
\email[Corresponding author; Email address: ] 
{joachim@thphy.uni-duesseldorf.de} 
\homepage{www2.thphy.uni-duesseldorf.de/~joachim}
\affiliation{Institut f{\"u}r Theoretische Physik II,
Heinrich-Heine-Universe{\"a}t D{\"u}sseldorf,
Universit{\"a}tsstra{\ss}e 1, D-40225 D{\"u}sseldorf, Germany}
\affiliation{Materials Research Laboratory, 
University of California, Santa Barbara, 
Santa Barbara, CA 93106, USA} 
\affiliation{Kavli Institute for Theoretical Physics, Kohn Hall, 
  University of California, Santa Barbara, CA 93106, USA}
\author{A.~G.~Moreira}
\affiliation{Materials Research Laboratory, 
University of California, Santa Barbara, 
Santa Barbara, CA 93106, USA} 
\author{P.~A.~Pincus}
\affiliation{Materials Research Laboratory, 
University of California, Santa Barbara, 
Santa Barbara, CA 93106, USA} 
\affiliation{Physics Dept., KAIST, 305-701 Daejon, Rep.\ of  Korea}

\date{\today, submitted to Macromolecules}


\begin{abstract}
We theoretically study the polarizability and the interactions of
neutral complexes consisting of a semi-flexible
polyelectrolyte adsorbed onto an oppositely charged spherical colloid. In the systems we studied,
the bending energy of the chain is small compared to the Coulomb
energy and the chains are always adsorbed on the colloid. We observe that 
the polarizability is large for short chains and small electrical fields and shows a non-monotonic behavior with
the chain length at fixed charge density. The polarizability has a maximum
for a chain length equal to half of the circumference of the colloid. For long chains we recover the 
polarizability of a classical conducting sphere. For short chains, the existence of a permanent dipole
moment of the complexes leads to a van der Waal's-type long-range attraction between them. This 
attractive interaction vanishes for long chains (i.e., larger than the colloidal size), 
where the permanent dipole moment is negligible.
For short distances the complexes interact with a deep short-ranged attraction which
is due to energetic bridging for short chains and entropic bridging for long chains. 
Exceeding a critical chain length eventually leads to a pure repulsion. 
This shows that the stabilization of colloidal suspensions by 
polyelectrolyte adsorption is strongly dependent on the chain size relative to the colloidal
size: for long chains the suspensions are always stable (only repulsive forces between the
particles), while for mid-sized and short chains there is attraction between the complexes and
a salting-out can occur.
\end{abstract}

\maketitle

\section{Introduction}

Polyelectrolyte(PE)--colloid complexes have recently motivated a great amount of
computational and analytical work
\cite{Kunze:prl:00,Sakaue:prl:01,Nguyen:jcp:01:2,gelbart:phystoday:00,%
Wallin:langmuir:96,Wallin:jcp:98,Netz:macromolecules:99,Mateescu:epl:99,%
Schiessel:epl:00,Nguyen:physicaa:00,Welch:macromolecules:00,Schiessel:jcp:01,%
Chodanowski:jcp:01,Jonsson:jcp:01:1,Jonsson:jcp:01:2,akinchina:macro:02,%
Gurovitch:prl:99,Nguyen:jcp:01:1,Park:epl:99,Granfeldt:jpc:91,Podgornik:jcp:95}.
 The reason for this is not only the
interesting behavior seen in such systems, but also their potential
applications. For instance, some water-soluble paints are formed by
PE--colloid complexes, in which a colloidal suspension  is stabilized by adsorbed
charged polymers. A better understanding of these complexes can give hints on
how to avoid the precipitation of the particles from the suspension.
Another example is the DNA-histone complex\cite{Kunze:prl:00,Sakaue:prl:01,Nguyen:jcp:01:2},
whose behavior is thought to be one of the crucial factors in the packing of DNA in living 
cells\cite{gelbart:phystoday:00}.

Most of the previous work on PE-colloid complexes looked essentially at the
structure and complexation behavior of the aggregates, e.g., of one PE and one macroion, 
\cite{Wallin:langmuir:96,Wallin:jcp:98,Netz:macromolecules:99,Mateescu:epl:99,
Schiessel:epl:00,Nguyen:physicaa:00,Kunze:prl:00,Welch:macromolecules:00,%
Schiessel:jcp:01,Chodanowski:jcp:01,Jonsson:jcp:01:1,Jonsson:jcp:01:2,%
akinchina:macro:02,Gurovitch:prl:99},
one PE and many macroions \cite{Jonsson:jcp:01:1,Jonsson:jcp:01:2,Nguyen:jcp:01:1,Schiessel:jcp:01},
one macroion and many PEs \cite{Wallin:jcp:98}, or the effect of overcharging of the macroion
\cite{Mateescu:epl:99,Park:epl:99,Nguyen:physicaa:00,Nguyen:jcp:01:1,Gurovitch:prl:99}.
While some works discuss the interactions between two macroions in
presence of either one or  more than two short PE-chains
\cite{Granfeldt:jpc:91,Podgornik:jcp:95,Nguyen:jcp:01:1}
we  concentrate here on the complex formed by one spherical colloid and one PE as a whole
and look at its polarizability, as well as the effective interaction
\cite{likos:physrep:01} between two isolated complexes. 
To do this we perform Monte Carlo (MC) computer simulations
of one and two neutral symmetric PE-colloid complexes, i.e., systems
where the charge of the PE exactly neutralizes the charge of the colloidal
particle. We focus on the case where the bending energy is small
compared to the energy of Coulomb coupling, i.e., the chain is
always completely adsorbed on the macroion.

While the results we will show are, strictly
speaking, only valid in the limit of low density of complexes and zero salt concentration,
we expect these to remain qualitatively correct under more realistic conditions. As we observe
in the simulations, the effect of salt at low concentrations (put into the simulations through a 
Debye-H{\"u}ckel interaction) affects only slightly the results given for the salt-free system.
Also, while both the polarizability and the interaction between complexes is 
very much influenced by the ``microscopics'' (i.e., by the structure of the chain),
we will show that most of the computer simulation results can be understood using
fairly simple principles.

The paper is organized as follows: in section~\ref{simulation} we describe the model used
and the computer simulation details. In section~\ref{polarizability} we present the simulation 
results for the dipole moment in an external electrical field of one PE-colloid complex and suggest
 analytical expressions for the 
polarizability for all chain lengths. Section~\ref{interaction} deals with the interaction of two 
of these complexes for different distances and chain lengths. We describe how the force is calculated and
the results are discussed. We provide analytical results for the force for short chain lengths and long 
distances. Finally in section~\ref{conclusion} we make some final remarks.

\section{The simulation model}
\label{simulation}

We performed monomer-resolved Monte Carlo (MC) simulations using
the model of Kremer {\it et al.}\ for single polyelectrolyte (PE) 
chains\cite{Stevens:kremer:jcp:95, Grest:87:1, Grest:94:1}.
The polyelectrolyte chains are modeled as bead-spring chains of $N$ Lennard-Jones
(LJ) particles. This method was first applied to neutral linear polymers 
and to a single star polymer \cite{Grest:87:1, Grest:94:1} later also for neutral 
\cite{Jusufi:macromolecules:99,Jusufi:jpcm:01} and charged stars polymer 
systems\cite{Jusufi:jcp:02,Jusufi:prl:02}.
For good solvent conditions, a shifted LJ potential is used to describe the purely
repulsive excluded volume interaction between all $N$ monomers:
\begin{eqnarray}  
  V_{\rm LJ}(r) = \left\{ 
    \begin{array}{ll}
      4\varepsilon_{\rm LJ}\left[
        \left(
          \frac{\sigma}{r}\right)^{12}
        -\left(\frac{\sigma}{r}\right) ^{6}+\frac{1}{4}
      \right] & \mbox{for $r\leq 2^{1/6}\sigma$}; \\
      0  & \mbox{for $r>2^{1/6}\sigma$}.
    \end{array}
  \right. 
  \label{vlj.eq}
\end{eqnarray}
Here, $r$ is the distance of the interacting beads, 
$\sigma$ is the microscopic length scale
of the beads and $\varepsilon_{\rm LJ}$ sets the energy scale. 
We will from now on rescale all lengths with $\sigma$ according to
\begin{equation}
  \til{r} = \frac{r}{\sigma}.
\end{equation}
In accordance 
with previous work\cite{Jusufi:macromolecules:99,Jusufi:jcp:02,Jusufi:prl:02},
we have chosen for the temperature 
$T = 1.2\varepsilon_{\rm LJ}/\kb$, where $\kb$ is the Boltzmann constant.

The connectivity of the bonded monomers is assured by a finite
extension nonlinear elastic (FENE) potential acting between neighboring beads:
\begin{eqnarray}  
  V_{\rm F}(r) = \left\{ 
    \begin{array}{ll}
      -\frac{1}{2}k_{\rm F}\left(\frac{r_{\rm m}}{\sigma}\right)^{2}
      \ln\left[1-\left(\frac{r}{r_{\rm m}}\right)^{2}\right]
      & \mbox{for $r\leq r_{\rm m}$};
      \\
      \infty 
      & \mbox{for $r>r_{\rm m}$} \, ,
    \end{array}
  \right.
  \label{vfene.eq}
\end{eqnarray}
where $k_{\rm F}$ denotes the spring constant and is set
to $k_{\rm F}= 7.0\varepsilon_{\rm LJ}$.
This interaction diverges at $r=r_{\rm m}$, which determines the maximal 
relative displacement of two neighboring beads. The energy 
$\varepsilon_{\rm LJ}$ is the same as in Eq.~(\ref{vlj.eq}), whereas for the 
length scale $r_{\rm m}$ we have chosen the value $\til{r}_{\rm m} = 2.0$.

Finally, the full Coulomb interaction $V_{\rm C}(r)$ 
between all charged monomers has to be taken into account: 
\begin{equation}
  \label{vclmb.eq}
  V_{\rm C}(r_{ij},Z_i,Z_j)=\frac{Z_i \, Z_j \, e^2}{4\pi\epsilon_{0}\epsilon_{r} r_{ij}} \equiv
  \kb T\lb\frac{z^2}{r_{ij}},
\end{equation}
where $Z_i$ and $Z_j$ are the number of elementary charges of the
interacting particles $i$ and $j$, 
$e$ is the elementary charge and $r_{ij}$ the distance between the particles. 
In our polymer model, every 
monomer has charge valence $z$, meaning that the total charge valence of one PE is simply
given by $Z \equiv N \, z $. The Bjerrum length $\lb$ is defined as the length at
which the electrostatic energy equals the thermal energy, viz.\
\begin{equation}
  \label{lb.eq}
  \lb = \frac{e^{2}}{4\pi\epsilon_{0}\epsilon_{r} \, \kb T},
\end{equation}
where $\epsilon_{0}\epsilon_{r}$ is the permitivity of the solvent. 
For water in room temperature one obtains $\lb=7.1{\rm \AA}$. 
The solvent is only taken into account via
the dielectric constant $\epsilon_{0}\epsilon_{r}$. In our simulations,
the Bjerrum length is fixed to $\lbt = 3.0$, which is a realistic value for typical 
polyelectrolytes, such as the hydrophobic sodium 
poly(styrene-{\it co}-styrene sulfonate) (NaPSS) or the hydrophilic 
poly(acrylamide-{\it co}-sodium-2-acrylamido-2-methylpropane-sulfonate).\cite{essafi:etal:95} 

The interaction between one PE and a spherical macroion of radius
$R_0$ is modeled as follows: a monomer in distance $r$ from the center of the colloid 
has a repulsive interaction  
$V_{\rm LJ}^{\rm m}(r)$
of the truncated and shifted Lennard-Jones type
\begin{eqnarray}
  V_{\rm LJ}^{\rm m}(r) =\left\{
    \begin{array}{ll}
      \infty & \mbox{for $r \leq R_0$};\\
      V_{\rm LJ}(r - R_0) & \mbox{for $r > R_0$},
    \end{array}
  \right.
  \label{mon.core1.eq}
\end{eqnarray}
and an attractive Coulomb interaction $V_{\rm C}(r,-Z,z)$, i.e.,  
electrostatic interaction between a monomer of charge valence $z$ and
an opposite charge at the center of the macroion with valence $Z = N \, z$,
which exactly neutralize the charge of the polyelectrolyte.
Naturally, the repulsive interaction between two macroions each with charge valence
$Z$ in distance $r$ is given by $V_{\rm C}(r,Z,Z)$. At a later stage we 
will include the effect of added salt by using a Debye-H{\"u}ckel
(screened) interaction, via 
\begin{eqnarray}
 V_{\rm DH}(r_{ij},Z_i,Z_j) = \lb\frac{Z_{i}Z_{j}}{\beta r_{ij}}\frac{\exp\left[-\kappa
 (r_{ij}-R_{i}-R_{j})\right]}{(1+\kappa R_{i})(1+\kappa R_{j})}
\end{eqnarray}
where $R_{i}$ and $R_{j}$ are the hard core radii of the two
interacting particles, in particular $R_{0}$ for the macroions and
zero for the monomers and $\kappa$ is the inverse screening length \cite{hansen:loewen:arpc:00}.
The inverse temperature is denoted by $\beta=1/\kb T$.
$\kappa$ is related to the salt molar concentration via
\begin{eqnarray}
c_{s}=\kappa^{2}/2N_{A}\lb,
\end{eqnarray}
where $N_{A}$ is Avogadro's number. We stress the fact that in the absence of salt, 
the system contains no small counterions: the chain and colloidal particle exactly 
cancel each others charge.

In this work, we focus our attention on the case where the 
bending energy of the chain is small compared to the Coulomb energy of
the complex. A good estimate of the  bending energy of a chain with persistence length
$l_{p}$ (isolated chain) and contour length $l$ wrapped around
a macroion to build a complex with radius $R$ is given by 
\begin{eqnarray}
E_{\rm bend}=k_{B}T \frac{l_{p}l}{2R^{2}}
\end{eqnarray}
This should be small compared to the Coulomb energy, what yields
\begin{eqnarray}
2Z^{2}\frac{\lb R}{l_{p}l}\gg1
\label{bending}
\end{eqnarray}
as the condition for a complete adsorption of the chain. We measured
the persistence length of an isolated chain by calculating the
bond vector correlations in dependence of the contour distance between 
the bonds and determining the typical decay length. We find a linear
dependence $l_{p}\approx l/2$ for a wide range of contour lengths if
no intrinsic stiffness is applied.
Inserting this scaling relation in (\ref{bending}) and using $Z \approx
\til l
z$ gives us a simple estimate for the wrapping state for chains with
no intrinsic stiffness
\begin{eqnarray}
4z^{2}\til{l}_{\rm B} \til R \gg 1.
\end{eqnarray}
Another limit where desorption can take place if the Coulomb coupling
is small compared to the thermal fluctuations, e.g., the macroion is
very big compared to the chain size. There is a complete adsorption of
the chain onto the colloid if the electrostatic attraction is  
larger than the thermal fluctuations
\begin{eqnarray}
   Z^2 \frac{\lb}{R} \gg 1.
  \label{adcond}
\end{eqnarray}
Strictly speaking, the adsorbed state is metastable: if the
colloid and the finite chain are initially put far apart from
each other, they interact essentially like two point-like charges.
It is well known that the $1/r$ potential between two
charges is not enough to bind them together, i.e., the Coulomb
potential between two point charges is not enough to compensate
for the loss of entropy through binding (cf.\ Manning condensation\cite{manning}).
However, if the particles are put enough close to each other, the system needs
a very long time before the unbound stable state is reached. To see this,
let us assume that two oppositely charged particles are confined to a ``spherical'' box of radius $L$. The 
probability of finding them at distance $r$ from each other is given by
\begin{equation}
  {\cal P}(r) = \frac{\exp\{Z^2 \lb / r \}}{\cal Z}
\end{equation}
where ${\cal Z} = \int_R^L dr {\cal P}(r)$ and $R$ is the contact distance between the charges. 
The ratio between the probability of finding the particles at $r'=r+\lb$ and at $r$ is then
given by
\begin{equation}
 R_{\cal P}  = \exp \{ - Z^2 \lb^2 / (r(r+\lb)) \} \ .  
\end{equation}
For $r=\lb$ and $Z=10$, this corresponds to $R_{\cal P} \sim 2 \times 10^{-22}$. In other
words, if the two charges are initially put $\lb$ apart, the probability of finding them at
$2 \lb$ apart is much smaller than to find them at $\lb$. Note that $R_{\cal P}$ increases
with $r$: if the particles are initially put very far apart from
each other (say $r/\lb$ of the order of $Z$), the system will flow to the stable
state much faster, since the ratio $R_{\cal P}$ become of order unity.
In our simulations, and in agreement with previous works, the PE has always stayed
at the close vicinity of the colloidal particle, since the parameters are chosen
such that Eq.~(\ref{bending}) and Eq.~(\ref{adcond}) are always satisfied.

Increasing chain stiffness has great influence on polarizability and 
interaction of these complexes. Some possible effects will be discussed briefly 
by introducing a simple harmonic coupling between the bonds of one PE
\begin{equation}
  V_{\rm bond}=\sum_{i=2}^{N-1} k_{\rm ang}(\alpha_i-\alpha_0)^2,
  \label{harm}
\end{equation}
where $\alpha_i$ is the angle formed between the connecting vectors of bonds $i$ and $i-1$ and 
the bonds $i$ and $i+1$. We choose an equilibrium angle $\alpha_0=\pi$ and examine the effects of increasing
chain stiffness by enhancing the bonding constant $k_{\rm ang}$. Unless if
explicitly mentioned, $k_{\rm ang}$ is always chosen to be zero.
 
We applied the standard MC Metropolis algorithm \cite{allen:tildesley}
 where one MC step consists of a trial move of every monomer.
The macroions were kept fixed. The acceptance rate for one MC sweep was approximately 50 percent. 
For faster equilibration and better statistics we introduced full chain moves where the whole PE is 
rotated around the closest macroion. Details describing these methods
can be found in \cite{Madras:statphys:88,joe:pre:00}. 
Our systems were simulated without any cell in an infinitely wide
 space where the full accurate Coulomb interaction without an cutoff 
is taken into account.  
We observed that only at low Coulomb coupling or large intrinsic stiffness
the thermal fluctuations lead to an entropic uncoupling of chain and colloid.
After a long equilibration time  
(typically of $1.5 \times 10^5$ to $ 2 \times 10^5$ MC steps), 
the different observables were averaged over runs with $10^6$ to $10^7$ MC steps.

\section{Polarizability of a single complex}
\label{polarizability}

\begin{figure} 
  \begin{center}
    \includegraphics[width=7cm,angle=-90]{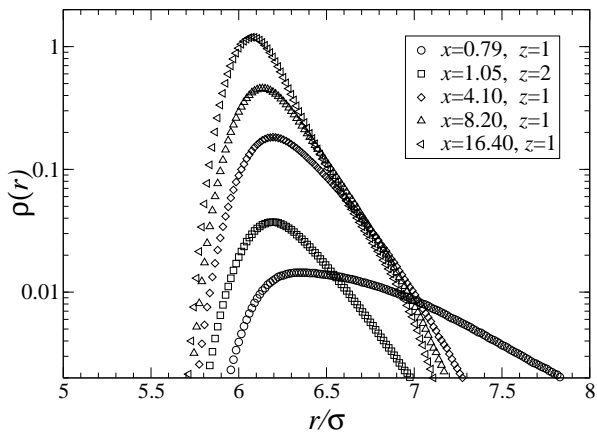} 
    \caption{Density profile of the monomers of one chain adsorbed to a spherical colloid with 
      radius $\til{R}_0=5$ for different chain length $x$ and different number of charges per 
      monomer $z$ plotted versus 
      the distance $\til{r}$ to the center of the colloid.
      With increasing $x$ and $z$ the monomers move closer to the center and the effective complex 
      radius $R$ decreases. See Fig. ~\ref{ss1} for simulation
      snapshots of one isolated complex.}
    \label{profiles}
  \end{center}
\end{figure}
In this section we study the polarizability of a single complex consisting
of one PE-chain with contour length $l$ and linear charge density $\tau=Ze/l$ ($Z= N \, z$
and $e$ is the elementary charge unit)
which is adsorbed onto a spherical colloid with radius $R_0$ and charge valence $Z$.
In order to calculate the contour length in the simulation, we measure the mean distance between
neighboring monomers: for a monomer charge $z=1$ we obtain $l\approx N\sigma$ and for $z=2$ we obtain 
$l\approx 1.3N\sigma$, i.e., $\tau\sigma=\til \tau=1$ for $z=1$ and
$\til \tau=1.54$ for $z=2$. In what follows we will extensively use as
a parameter the ratio between the PE length and the \emph{effective}
diameter of the complex ($2 \, R$) 
\begin{equation}
  x \equiv \frac{l}{2R}.
\end{equation}

\begin{figure} 
  \begin{center}
    \includegraphics[width=8cm, angle=0]{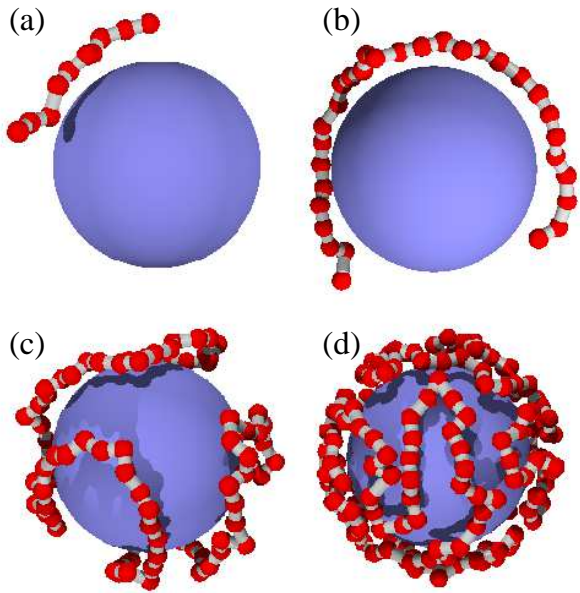}
    \caption{Computer simulation snapshots of a PE adsorbed to a spherical colloid 
      with radius $\til{R}_0=5$ for (a) $x= 0.79$, (b) $x=2.78$, (c) $x=8.20$ and (d) $x=16.40$. 
      The number of elementary charges per monomer is $z=1$. The
      monomers are rendered as dark spheres with diameter $\sigma$.} 
    \label{ss1}
  \end{center}
\end{figure}

For $x=\pi$ the chain length is equal to the
circumference of the complex. For a chain much thinner than the
colloid radius and at low temperatures, the effective
radius $R$ should be given by $R\approx R_0$. However, since in our simulations the 
monomers have a size that is comparable to the colloidal particle,
this effective radius is approximately $R \approx R_0+\sigma$, with some dependence
on the number of monomers in a chain and on the charge of each monomer, 
as can be seen in Fig.~\ref{profiles}.
In this figure, we show the density profile $\rho(r)$ of the monomers surrounding a colloid 
with radius $\til{R}_0=5$ for chains with different number of monomers $N$ and 
different monomer charges $z=1,2$. As one would expect, a higher Coulomb coupling between
the PE and the colloid (through a higher total charge in the PE) leads to a reduction
of the effective radius of the complex, here assumed to be at the maximum of the
monomer distribution. Simulation snapshots are shown in Fig.~\ref{ss1} 
for systems with different numbers of monomers and no external field. As already
mentioned, the dependence of $R$ on the parameters of the PE is so weak that it
can be assumed for all practical purposes to be a constant 
for a wide range of chain lengths. In the following we will choose
$\til R=6.3$ for short chains $x< \pi$ and $\til R=6.1$ for
long chains $x>\pi$ as effective complex radius, for a fixed
colloid radius $\til R_{0}=5.0$ and $z=1$. See Tab. \ref{tab} for a
summary of the relation between complex radius, monomer number, and
chain length for the simulated system with $z=1$. The lengths for
$z=2$ can be calculated accordingly; here we used $R=6.2$ for $x<\pi$
and $R=6.0$ for $x>\pi$. The colloid valence is simply given by
$Z=Nz$.
We checked that all forthcoming results obtained from the
simulations are independent of above particularly chosen radius of the colloid. All
theoretical results in this and the following section should hold
generally for all colloid sizes.

\begin{table}
\caption{\label{tab}Parameters of the simulated system for $z=1$ and
  $\til R_{0}=5$. We 
  simulated chains with different monomer numbers $N$ and chose a
  fixed complex radius $R$ according to the density profiles, see Fig.
  \ref{profiles}. $x=l/2R$ is the according chain lengths, while
  $l=N\sigma$ for $z=1$.}
\begin{ruledtabular}
\begin{tabular}{l l l l l l l l l l l l}

$N$  & 10 & 15 & 20 & 25 & 30 & 35 & 40 & 50 & 100 & 200 & 300 \\
\hline
$\til R$  & 6.3 & 6.3 & 6.3 & 6.3 & 6.3 & 6.3 & 6.1 & 6.1 &
6.1 & 6.1 & 6.1\\ \hline
$x$  & 0.79 & 1.19 & 1.59 & 1.98 & 2.38 & 2.79 & 3.27 & 4.10 &
8.20 & 16.40 & 24.60\\
\end{tabular}
\end{ruledtabular}
\end{table}
We start our analysis with chains which are short compared to the size of the colloid,
viz.\  $x < \pi$. In this regime the PE-colloid complex has always a nonvanishing
instantaneous dipole moment, since the center of charge of the chain cannot 
coincide with the center of the colloid. We define the dipole moment $\vec{P}$ 
of the complex as 
\begin{eqnarray}
  \vec{P}=\frac{Z e}{N}\sum_{i=1}^N (\vec{r_i}-\vec{R}_c),
\end{eqnarray}
where $\vec{R}_c$ is the vector pointing to the center of the colloid and $\vec r_i$ is the 
position of monomer $i$. Notice that, with this definition, the magnitude of $\vec{P}$ is
given by $Ze$ times the distance between the center of the colloid and the center of charge
of the PE-chain. It is easy to show that the mean dipole moment of a thermally fluctuating rigid 
dipole with length $d$ and charge $Z e$ in an electrical field $E$ is 
\begin{eqnarray}
  P = Z e \, d\left({\rm coth}(\beta \, Z e \, d \, E)-\frac{1}{\beta \, Ze \, d \, E}\right),
  \label{dipole_old} 
\end{eqnarray}
where $P$ is the magnitude of $\vec{P}$ and $\beta \equiv 1/k_B T$. 
Introducing the rescaled dimensionless dipole moment
\begin{equation}
  P^*=\frac{P}{\sigma e }
\end{equation}
and electric field
\begin{equation}
  E^*=\beta \sigma e \, E
\end{equation}
one can rewrite Eq.~(\ref{dipole_old}) as
\begin{eqnarray}
  P^*=Z \, \til{d}
  \left({\rm coth}(Z \, \til{d} \, E^*)-\frac{1}{Z \, \til{d} \, E^*}\right),
  \label{dipole} 
\end{eqnarray}
where $\til{d}=d / \sigma$. For low temperatures or strong electrical 
fields $ Z \til{d}  E^*  \gg 1$, the mean dipole moment saturates to $P^* = Z \til{d}$, 
while for small fields $ Z \til{d} E^* \ll 1$ and we recover the well known
result 
\begin{eqnarray}
  P^*=\frac{ Z^2 \til{d}^2}{3} \, E^*.
  \label{smallE}
\end{eqnarray}
For the case of a chain adsorbed onto a spherical colloid, the center of charge of the chain
is on average somewhere between the center of the colloid and its surface. Assuming that the chain 
can be approximately described by a circular arc fluctuating on the surface of the 
colloid (see eg.\ Fig.~\ref{ss1}(b)), the distance between the center of the colloid
and the center of charge of the chain is given by
\begin{eqnarray}
  \til{d}=\til{R}\,\,\frac{\sin (x)}{x},
  \label{R_eff}
\end{eqnarray}
where $\til{R}=R/\sigma$. Our model is assumed to work for chains
shorter than the circumference of the sphere, i.e Eq.~(\ref{R_eff})
holds for $0<x<\pi$.

\begin{figure} 
  \begin{center}
    \includegraphics[width=4.8cm, angle=-90]{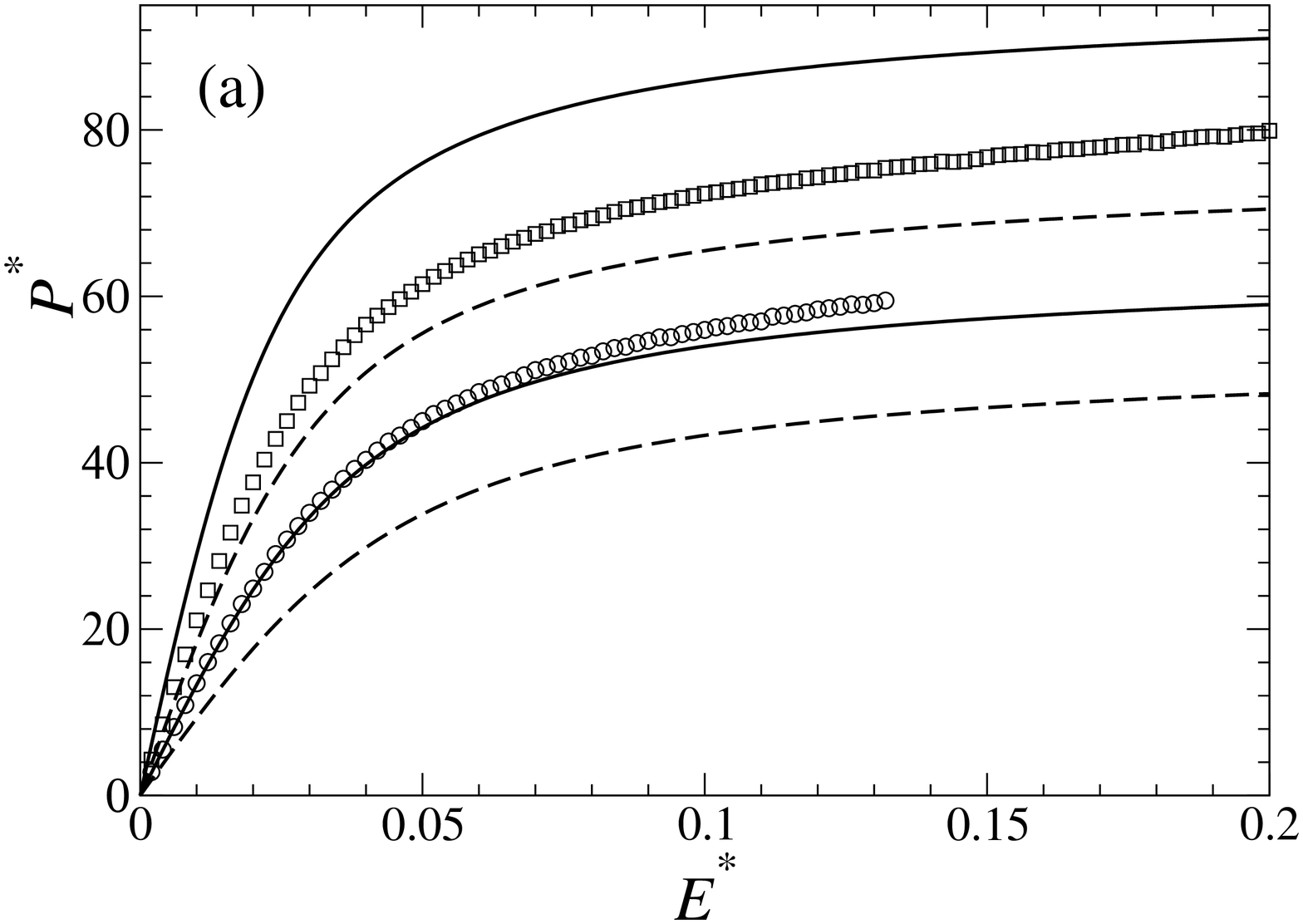}
    \includegraphics[width=4.8cm, angle=-90]{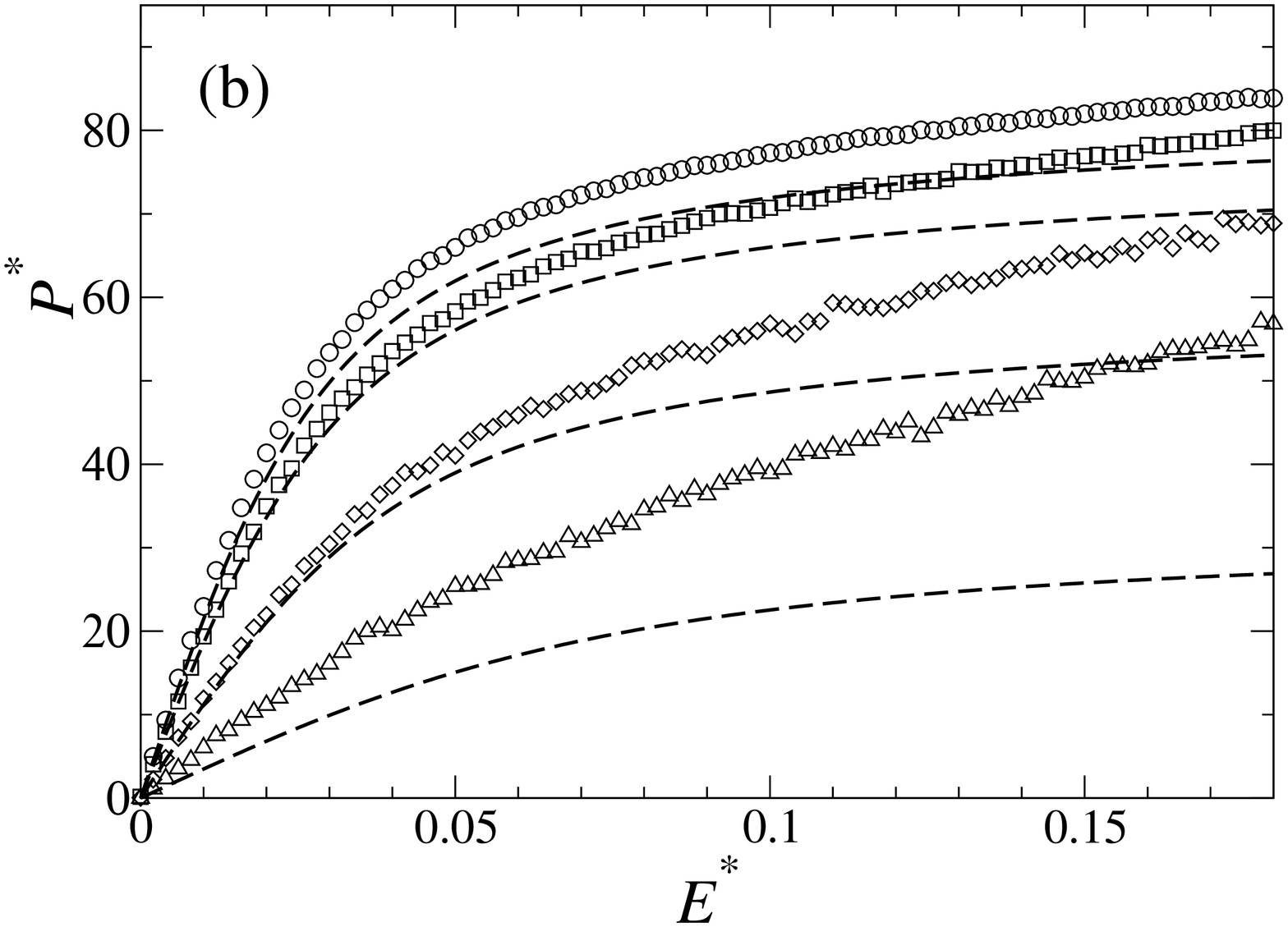}
    \includegraphics[width=4.8cm, angle=-90]{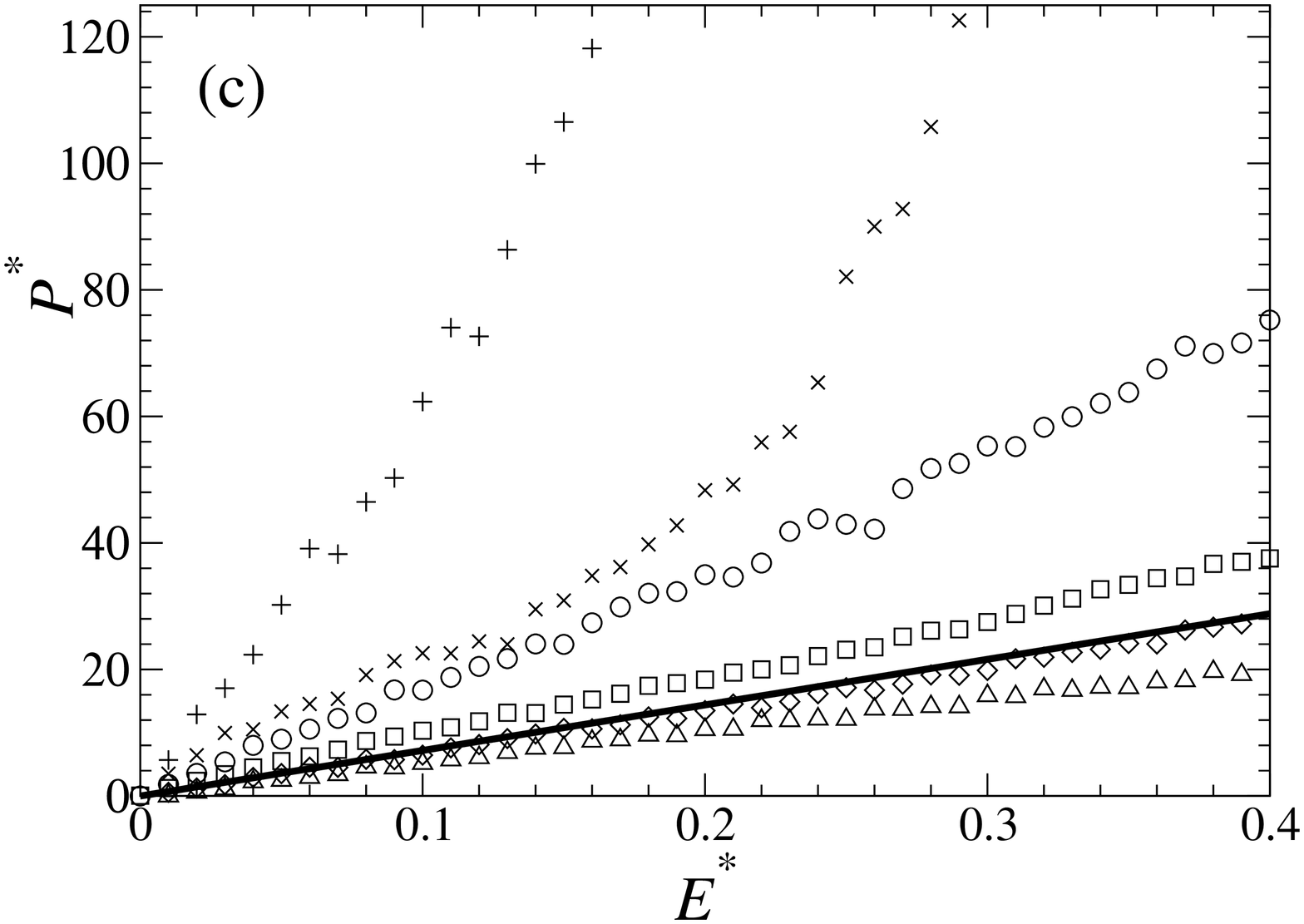}
    \caption{Dimensionless dipole moment $P^*=P/\sigma e$ of one PE-colloid complex versus the dimensionless 
      external electrical field $E^*=\beta \sigma e \, E$ 
      for different chain lengths, $z=1$  and a colloidal radius
      $\til{R}_0=5$. Symbols are computer simulation results, lines
      are theoretical results. Plot (a) shows the scaled chain lengths
      $x=0.79$ (circles) and $x=1.19$ (squares). The dipole 
      moment increases with $x$. Lines are the theoretical
      results, Eq.~(\ref{dipole}), using  $\til{d}=\til{R}$ (solid) or
      the effective dipole length $\til{d}=\til{R}\sin(x)/x$ (long dashed).
      Plot (b) shows the results for $x=1.59$ (circles),$x=1.98$ (squares), $x=2.38$ (diamonds), 
      $x=2.78$ (triangles). Note that the dipole moment now decreases
      with increasing $x$. The long dashed lines is theory as used in (a).
      Plot (c) shows the results for $x=4.10$(circles), $x=8.20$ (squares), $x=16.40$ (diamonds), 
      $x=24.60$ (triangles). The solid line is the dipole moment of 
      a conducting sphere with radius $\til{R}_{cs}=6.1$ according to $P^*=\alpha_{cs} E^*$,
      where $\alpha_{cs}$ is given by Eq.(\ref{sphere}). In (c) we
      also show the effect of increasing chain stiffness for $x=4.10$.
      The harmonic coupling is quantified by $\beta k_{\rm ang}=4.0$ (crosses)
      and  $\beta k_{\rm ang}=5.0$ (plusses).}
  \label{dipolemoment}
 \end{center}
\end{figure}

In Fig.~\ref{dipolemoment} we compare the dipole moment of 
a complex in an external field for various chain lengths and $z=1$ as a function of the
external field from computer simulations and from Eq.~(\ref{dipole}). Note that
$\til{R}_0=5$ and the effective radius is assumed to be $\til{R}=6.3$
for $x\lesssim \pi$. For the theoretical
effective dipole length $d$, we use the result from Eq.~(\ref{R_eff}) when
$0 \lesssim x \lesssim \pi$ and Eq. \ref{bending} is satisfied.
If  Eq. \ref{bending} is not fulfilled, the chain is essentially
a straight rod that touches the sphere---in this case, 
the effective length of the dipole is simply given by
$R$. On the other hand, for $x \gtrsim \pi$ the chain cannot be approximated by
a two-dimensional circle and Eq.~(\ref{dipole}) breaks down. 

\begin{figure} 
  \begin{center}
    \includegraphics[width=7cm,
    angle=-90]{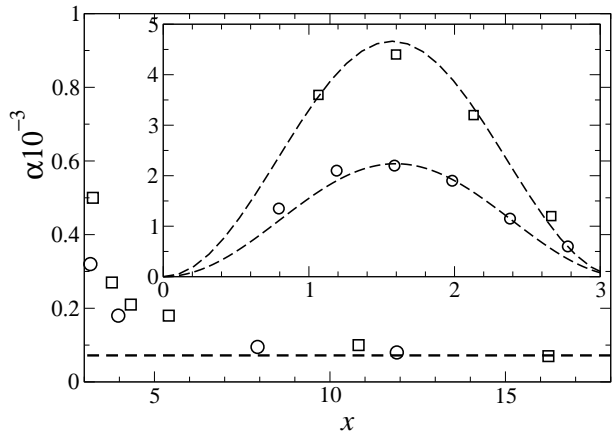} 
    \caption{Dimensionless polarizability $\alpha$ versus the scaled chain length $x$
      at small applied fields. The symbols denote the simulation, the dashed lines are 
      the theoretical results from Eq.~(\ref{polar.eq}) (inset) and
      Eq.~(\ref{sphere}) for charges per monomer $z=1$
      (circles) and $z=2$ (squares). The colloid radius is $\til R_{0}=5$. }
    \label{alpha_x}
  \end{center}
\end{figure}

The polarizability $\alpha$ of the complex is defined as the slope of its dipole moment
as a function of the electric field
\begin{eqnarray}
\alpha = \frac{dP^*}{dE^*} \, .
\end{eqnarray}
The behavior of $\alpha$ is shown in
Fig.~\ref{alpha_x}: in the regime of small electric fields, the
polarizability is constant, here we are in the linear response regime,
see Eq.~(\ref{smallE}).

In the regime where the interaction energy between the dipole and the applied field
largely exceeds the thermal energy ($ Z \til{d}  E^* \gg 1$) 
the polarizability becomes very small; this latter behavior is not
surprising, since in this regime the dipole is essentially pointing towards the field,
and thermal fluctuations which tend to destroy this alignment at small fields become 
unimportant. In our simulations we never really reach the point where the polarizability
becomes zero since at some point the field is strong enough to induce the unbinding of
the chain from the colloidal particle.
Plot \ref{dipolemoment}(a) shows the dipole moment for two short chains, viz.\ $x=0.79$ and $x=1.19$.
Notice that the system with former value of $x$ is very well described by 
$P$ with $d=R$, while the latter tends to approach the dipole moment which uses $d$
given by Eq.~(\ref{R_eff}). The reason for this is the fact that in the system with smaller
$x$ the polyelectrolyte has a bending energy of the order of the
Coulomb energy of the complex, see Eq. (\ref{bending}), and the system behaves more 
like a rigid rod in contact with a sphere than as a circular arc, as described 
by Eq.~(\ref{R_eff}). The curve for $x=0.79$ in (a) stops abruptly
since the critical field for the desorption is reached.  

\begin{figure} 
  \begin{center}
   \includegraphics[width=7cm, angle=0]{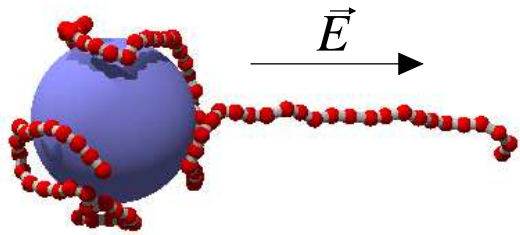}
    \caption{Simulation snapshot of a PE-colloid complex in an external electrical field $E$. 
      The chain has an intrinsic stiffness due to 
      the harmonic coupling Eq.~(\ref{harm}). The bonding constant 
      is $\beta k_{\rm ang}=4.0$, the number of monomers is $N=100$, 
      $E^*=0.4$, $z=1$,  and $\til{R}_0=5.0$.
      } 
    \label{ssE}
 \end{center}
\end{figure}
Longer chains lead to a decrease of the polarizability at small 
fields as can be seen in plot (b) of
Fig. \ref{dipolemoment}, where we plot $x=1.59$, 1.98, 2.38 and 2.78. This effect
is a result of the dislocation of the center of charge of the chain to a position
that is closer to the center of the colloid, according to Eq.~(\ref{R_eff}). 
Note that complexes with medium-sized chains tend to approximately exhibit
dipole moment of a circular arc on a sphere. However, complexes with
long chains, $x \gtrsim \pi$ tend to have a linear 
response to a wider range of the applied field (see plots (b) and (c)),
which is a consequence of the fact that larger chains 
have more charge at fixed charge density, and so the fields needed to 
deviate from a linear law are larger; also, the fact that longer chains have more
possible conformations leads to a qualitative change in the response of the
complex, and a simple model as the one used above is no longer valid.

Let us have a closer look at the polarizability for small fields and
short chains. Inserting Eq.~(\ref{R_eff}) into (\ref{smallE}), one
obtains, for small fields, the dimensionless polarizability of the
complex as a  function of the PE-chain length and charge density 
\begin{eqnarray}
  \alpha=\frac{4 \, \til{\tau}^2 \, \til{R}^4}{3}\sin^2 (x).
  \label{polar.eq}
\end{eqnarray}
where the linear charge density $\tau$ has been rescaled such 
that $\tilde{\tau}=\sigma \tau$. For a fixed $\tau$, this polarizability
shows a maximum at $x=\pi / 2$  for a chain with length equal to half the
circumference of the colloid and vanishes for  $x=\pi$, when a chain has a full
circumference. Indeed, this behavior predicted by theory 
is very close to in the computer simulation results in the inset of
Fig.~\ref{alpha_x}, 
where we plot $\alpha$ for small fields for a colloid with fixed
radius $\til{R}_0=5$ and $\til{R}=6.3$ and monomer charges $z=1$ and $z=2$.  

In the cases where the complex contains a long chain
$x \gtrsim \pi$, the polarizability Eq.~(\ref{polar.eq}) loses its validity.
Note that in this case, the chain is wrapped around the sphere more than one 
time and one expects a small or vanishing dipole moment.
Because the monomers tend to distribute themselves on average more or less homogeneously
throughout the surface of the sphere, one could expect the response of the PE-colloid
complex to electric fields to become similar to the response of a neutral conducting sphere,
which has a dimensionless polarizability given by (see. e.g.,
\cite{Jackson})
\begin{equation}
  \alpha_{cs} = \frac{\til{R}_{cs}^3 }{\lbt} 
  \label{sphere}  
\end{equation}
where $R_{cs}$ is the radius of the sphere and $\lbt = \lb / \sigma$.
Not surprisingly, in the limit of very large chains $x \gg \pi$ this latter result
is recovered, as shown in Fig.~\ref{alpha_x}. For this comparison we
have chosen $\tilde R_{cs}=6.1$ for the effective sphere radius.
Note that the simulation results for long chains are independent of
the  linear charge density in agreement with Eq. (\ref{sphere}).

Let us now discuss possible effects of an increasing 
intrinsic stiffness of the chain. For short chains we observe an increasing polarizability
if the harmonic coupling is strong enough to decrease the curvature of
the chain around the spherical macroion meaning that the bending
energy reaches values of the order of the Coulomb energy..  

The center of charge of the PE is less close to the
colloid center and the effective dipole length increases. 
As an upper limit the maximal dipole moment is then given by
Eq.~(\ref{dipolemoment}) inserting $d=R$, with $R$ effective complex radius. 
However, for the case, where the bending energy comes close to the
Coulomb energy, we observe that the chain decouple easily from the
colloid and no bound complex is formed. 

More interesting is the behavior for long chains. For long chains with
a large bending energy and no electrical
field, so called one-tail or two-tail configurations sets 
in\cite{Wallin:langmuir:96,Kunze:prl:00,Jonsson:jcp:01:2,Chodanowski:jcp:01,akinchina:macro:02}. 
Here part of the PE is still adsorbed on the colloid while one 
or two rigid parts are stretched away from the macroion to
lower the bending energy and so minimize the total energy. In an
electrical field the one tail 
configuration is obviously promoted due to the strong dipolar
interaction with the external field, see Fig. ~\ref{ssE} for a
simulation snapshot. We make the following observation in our
simulation: if the chain is fully adsorbed on the colloid the
polarizability is still given by Eq.~(\ref{sphere}) independent of
the higher chain stiffness, only the effective complex radius increases
slightly due to a less tight wrapping around the colloid.
But for a sufficiently strong chain rigidity a small external
electrical field can now pull a part of the chain in direction of the
chain to lower the bending and dipole energy and a dipolar one tail
configuration is formed. Two examples of the dipole moment
are shown in Fig. \ref{dipolemoment} (c) for $x=8.20$,  
$\beta k_{\rm ang}=4.0$ and $\beta k_{\rm ang}=5.0$. For small fields the chain is fully adsorbed and the 
polarizability is the same as for the $\beta k_{\rm ang}=0$ case. Then the increased stiffness leads to a
continuous dewrapping in the one-tail configuration and the
polarizability increases dramatically due to formed dipole.  
Further increasing of the electrical field lengthens 
the tail and eventually decoils the chain from the colloid. The longer
the tail, the larger is the dipole moment and the polarizability
increases with growing field until complete desorption is accomplished.
A snapshot for an one-tail complex in an electrical field $E$ 
is shown in Fig.~\ref{ssE}. 

Finally, adding salt obviously increases the polarizability as the range and strength of
the Coulomb attraction between PE and the colloid is screened and the
effective complex radius grows. We take the presence of salt into account by
using a Debye-H{\"u}ckel potential between the monomers (and ignoring any polarizability
due to the salt), going up to a salt concentration of $c_{s}\approx 1.0 {\rm M}$.
We observe that Eq.~(\ref{polar.eq}) for short
chains and Eq.~(\ref{sphere}) for long chains still hold as long a
stable complex is formed but with a slightly larger effective radius $R$ or
$R_{cs}$, respectively.

\section{Effective interaction between two complexes}
\label{interaction}

\begin{figure} 
  \begin{center}
    \includegraphics[width=6cm, angle=-90]{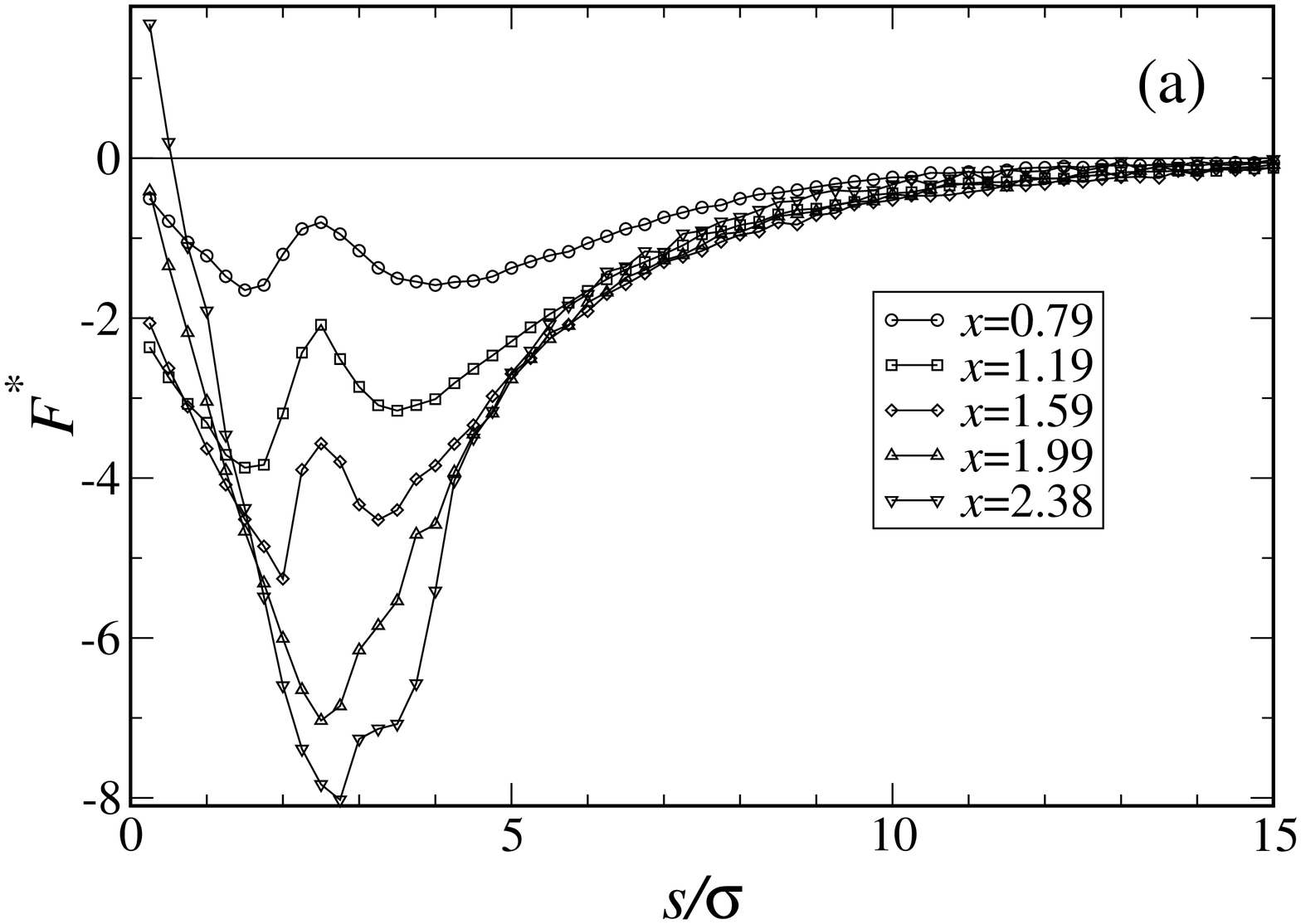} 
    \includegraphics[width=6cm, angle=-90]{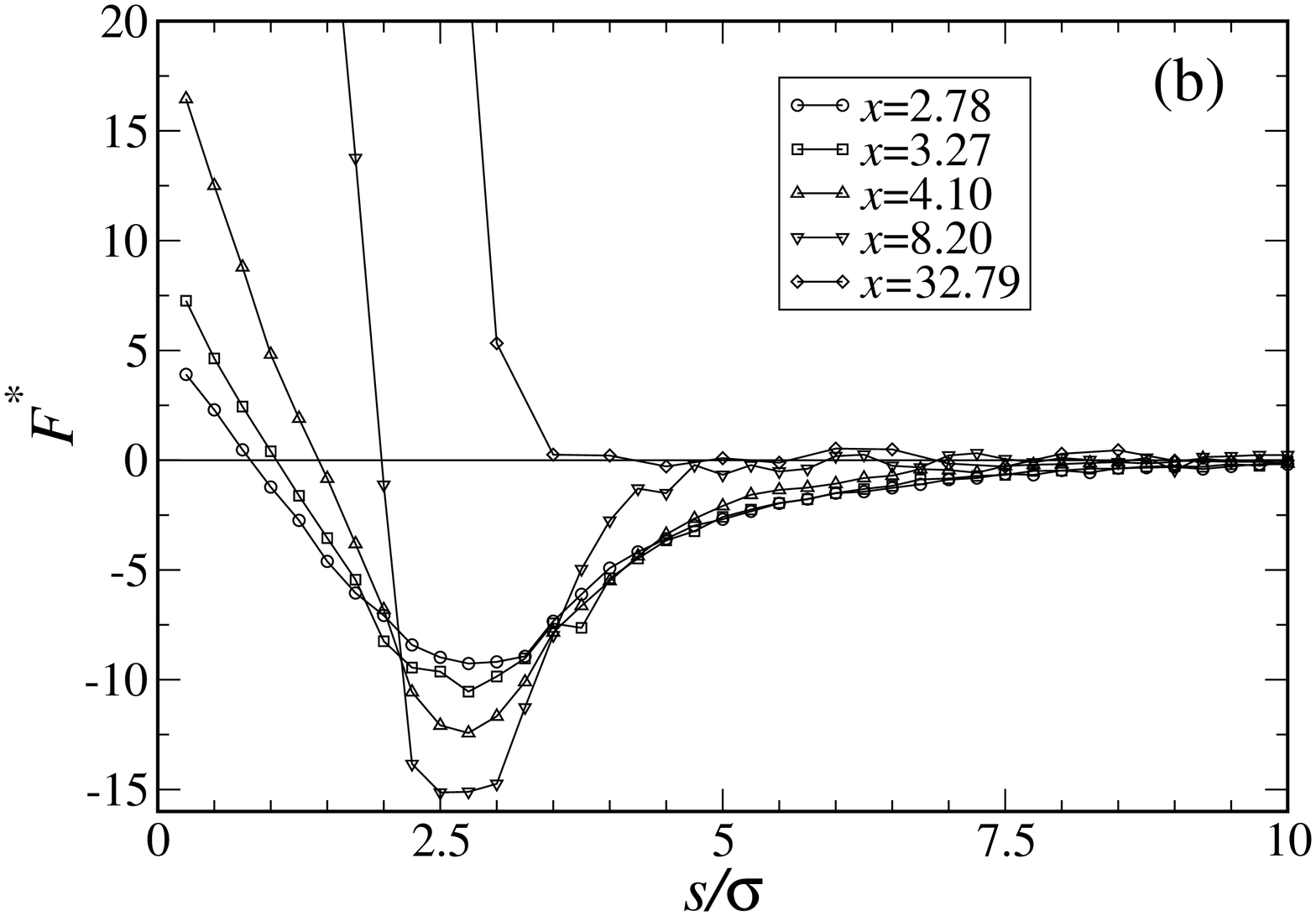} 
    \caption{Simulation results of the dimensionless force $F^*=\beta \, \sigma \, F$ 
      between two PE-colloid complexes versus the 
      surface-surface distance $\til{s}$ for different chain lengths
      $x$ and $z=1$. The colloid radius is $\til{R}_0=5.0$. Lines are guide to
      the eye.
      Plot (a): short chains $x\lesssim \pi$. 
      Plot (b): long chains $x\gtrsim \pi$. .
      }
    \label{force}
  \end{center}
\end{figure}

\begin{figure} 
  \begin{center} 
    \includegraphics[width=5.0cm, angle=0]{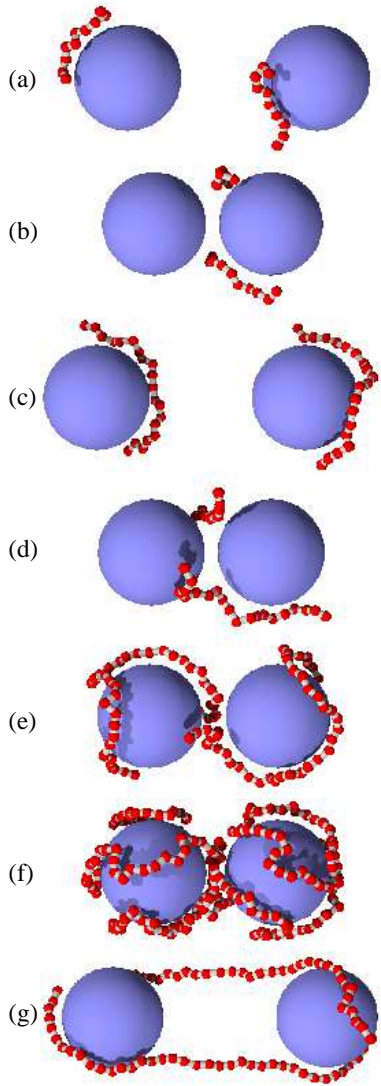}
    \caption{Snapshots of two interacting PE-colloid complexes 
      with colloid radius $\til{R}_0=5$, $z=1$, chain length $x$ and surface-to-surface distance $s$. 
      From top to bottom: (a) $x=0.79$, $\til{s}=8.0$; (b) $x=0.79$, $\til{s}=1.5$; (c) $x=1.59$, 
      $\til{s}=10.0$; (d) $x=1.59$, $\til{s}=1.5$; (e) $x=4.10$, $\til{s}=2.5$; (f) $x=8.20$, 
      $\til{s}=2.0$; (g) $x=3.27$, $\til{s}=11.25$, $\beta k_{\rm ang}=4.0$}
  \label{ss2}
 \end{center}
\end{figure}

We now turn to the interaction between two complexes at
distance $r$. The effective force acting on the center of one colloid in the 
simulation is calculated as follows:
consider two spherical colloids with radius $R_0$ and center-to-center distance $r$, 
which means that they have a surface-to-surface distance of $s= r - 2 R_0$. 
The energy of one colloid in presence of one monomer at distance $r^m$ is
the sum of Coulomb attraction and short-ranged LJ-repulsion
\begin{equation}
V_{\rm mon}(r^m)=V_{\rm C}(r^m,-Z,z)+V^m_{\rm LJ}(r^m).
\end{equation}
After a sufficiently long equilibration time, the force ${F}_m$ of all 
monomers acting on one of the colloids is measured by performing the 
statistical average
\begin{equation}
  {F}^m(r) =\left\langle \sum_{i=1}^M\sum_{j=1}^{N}
    \nabla_r V_{\rm {mon}}(r^m_{ij})
  \right\rangle,
  \label{f1.eq}
\end{equation}
where the sum in $i$ goes over all $M$ PE-chains and the sum in $j$ considers all $N$ monomers of 
each chain. $r^m_{ij}$ is the distance of monomer $j$ from the $i$th PE-chain to the center of
the considered colloid. The brackets in Eq.~(\ref{f1.eq}) denote ensemble statistical average.
We obtain the total force between the colloids in distance $r$ by adding the direct repulsive 
Coulomb force between the colloidal particles, viz. $Z^2 e^2 / 4 \pi \epsilon_{0}\epsilon_{r} r^2$.
The total force can be then divided into two contributions,
\begin{equation}
  F(r) = F_{\rm C}(r) + F_{\rm LJ}(r).
\end{equation}
where 
\begin{equation}
  F_{\rm C}(r)=\frac{Z^2 e^2 }{4 \pi \epsilon r^2}+\left\langle \sum_{i=1}^M\sum_{j=1}^{N}
    \nabla_r V_{\rm {C}}(r^m_{ij},Z,z)
  \right\rangle
  \label{force.coul}
\end{equation}
is the Coulomb force due to the electrostatic attraction of all  monomers and the electrostatic 
repulsion of the second colloid and 
\begin{equation}
  F_{\rm LJ}(r)=\left\langle \sum_{i=1}^M\sum_{j=1}^{N}
    \nabla_r V_{\rm {LJ}}(r^m_{ij})
  \right\rangle
  \label{force.lj}
\end{equation}
resulting from the excluded volume interaction of the monomers through the
aforementioned Lennard-Jones potential. Since this force is essentially
a hard-core exclusion force, one can consider this as describing the interaction due to 
gain and loss of entropy. For this reason we call it entropic force.

Based on the previous discussion on the polarizability of a single complex, one can essentially
expect two distinct regimes for the interaction between two equal symmetric complexes, 
viz.\ when the chains are short and 
when the chains are long. Fig.~\ref{force} summarizes our results, where the dimensionless
force $F^* = \beta \, \sigma \, F$ is plotted for systematical
increase of $x$ and $z=1$ as a function of $s$, the 
surface-to-surface distance. Negative forces mean attraction while positive
forces mean repulsion. See Fig.~\ref{ss2} for some typical simulation snapshots.
Notice that for short chains, a long-range attractive tail is present.
This is consistent, as we will demonstrate, with a dipolar interaction between the complexes. 
For longer chains, this long range tail disappears, and only a strong short-range attraction (also
present for shorter chains) survives. In the limit of very long
chains, only a repulsive steric interaction is present. We discuss
these results in more detail in the following. 

\subsection{Short chains}

\begin{figure} 
  \begin{center}
    \includegraphics[width=6cm, angle=-90]{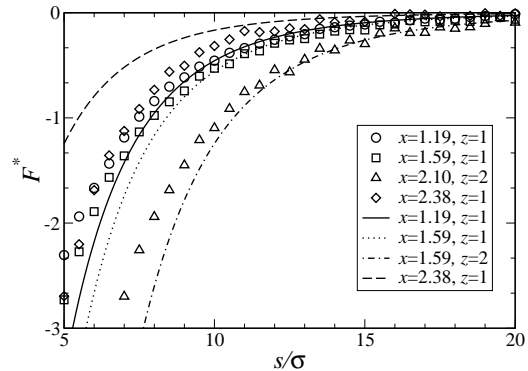} 
    \caption{Interaction force between two PE-colloid complexes for long distances and short chains
      ($x \lesssim \pi$). The plot shows a comparison made between the analytical
      expression Eq.~(\ref{force.eq}) for the dipole-dipole interaction (lines)
      and the computer simulation results (symbols) for colloids with size 
      $\til{R}_0=5$ and various chains lengths $x$ and monomer charge
      $z$.
      }
    \label{force_dipole}
  \end{center}
\end{figure}

In the previous section we showed that the complexes with short chains
respond like permanent dipoles to electric fields. When interacting with
each other, it is reasonable to expect that in such case a dipole-dipole
interaction shows up, at least at large separations between complexes.  
One way of estimating this long-range tail is by looking at the
interaction between two fluctuating permanent dipoles $P=Zed$ with length
$d$ as given by Eq.~(\ref{R_eff}) expressed through the so called Keesom
energy \cite{Keesom}
\begin{eqnarray}
\beta v(r)=\frac{1}{3}\frac{Z^{4}\til{l}_{\rm B}^{2} \til d^{4}}{\til
  r^{6}}.
\end{eqnarray}
After some algebra one obtains
for the interaction energy between two PE-colloid complexes with
$0\lesssim x\lesssim\pi$
\begin{equation}
  \beta v(r)=-\frac{16 \, \til{l}_{\rm B}^2 \, \til{\tau}^4 \, \til{R}^8}{3 \, \til{r}^6} \sin^4(x).
\end{equation}
where $r$ is the distance between the dipole and the polarizable object.
Accordingly, the attractive force acting on the polarizable object
is
\begin{equation}
  F^{*}(r)=-\frac{32 \, \til{l}_{\rm B}^2 \, \til{\tau}^4 \, \til{R}^8}{ \til{r}^7} \sin^4(x).
  \label{force.eq}
\end{equation}
We easily see that the electrostatic interaction has a maximum of
strength for a length $x=\pi/2$ according to the behavior of the
polarizability of one isolated maximum. Another remarkable feature is the
strong dependence on the complex radius which goes with the eighth power for
a fixed $x$. 
A comparison of expression  (\ref{force.eq})
with the long distance tail of the force between the complexes
is done in Fig.~\ref{force_dipole} for different chain lengths and
charge densities, leading as expected to a nice agreement. As
theoretically predicted the simulations show a maximal dipole
interaction for a length $x\approx \pi/2$.

\begin{figure} 
  \begin{center}
    \includegraphics[width=6cm, angle=-90]{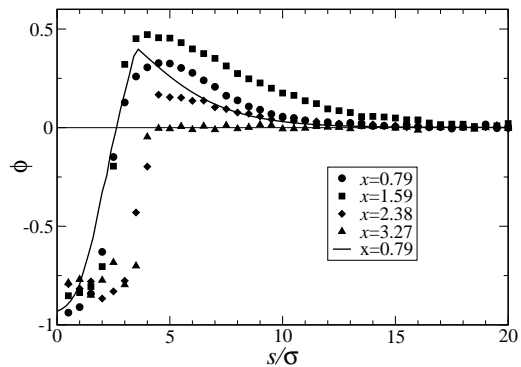} 
    \caption{Simulation results (symbols) of the order parameter $\phi$
      versus the surface-to-surface distance of two PE-colloid
      complexes with  $\til{R}_0=5.0$, $z=1$, and different chain lengths. The
      solid line is the numerical solution of the toy model sketched
      in Fig.~\ref{sketch} for $x=0.79$.}
    \label{phi}
  \end{center}
\end{figure}

In order to obtain some insight into the conformational properties of the
interacting complexes we introduce the order parameter
\begin{eqnarray}
  \phi=\frac{\vec P_1\vec P_2}{|\vec P_1||\vec P_2|} \, ,
\end{eqnarray}
which is the normalized dot product between the two dipole moments of each complex.
For a dipole-dipole interaction $-1\leq\phi\leq1$ should be positive as both dipoles
show in the same direction. Negative values of $\phi$ mean that the dipoles are on average
pointing in opposite directions. By studying the
average distance between the center of charge of the chains 
in our simulations, we found that every time 
$\phi$ becomes negative the monomers accumulate between the colloids,
meaning that each dipole moment points in  direction of the other
complex  in average.

Fig.~\ref{phi} shows $\phi$ as a function of separation 
for different chain lengths with $z=1$ at fixed complex radius. As expected, for short chains 
and large distances, $\phi$ is positive: the dipoles of the two complexes basically
point in the same direction and produce the dipole-dipole interaction as discussed
above. Note that the largest values for $\phi$ are obtained for complexes with 
chains with $x=1.59$, a value close to $x=\pi/2$ where the
polarizability is maximal. Typical configurations in the dipole-dipole
regime are shown in Fig. \ref{ss2} (a) and (c) where the order
parameter $\phi$ has positive values. A relatively sharp transition from positive to 
negative values of $\phi$ is observed for distances $\til s \approx 4$ for all values of $x$.

For small distances, $\phi$ becomes negative for all systems studied. As mentioned before,
this means that the centers of charge of the chains are located 
somewhere between the colloids, something that turns out to be always true 
as $\til s\lesssim 4$. This can also be seen in snapshots for close distances of the complexes 
as for instance in Figs.~\ref{ss2}(b) and (d).
 
\begin{figure} 
  \begin{center}
    \includegraphics[width=7cm, angle=0]{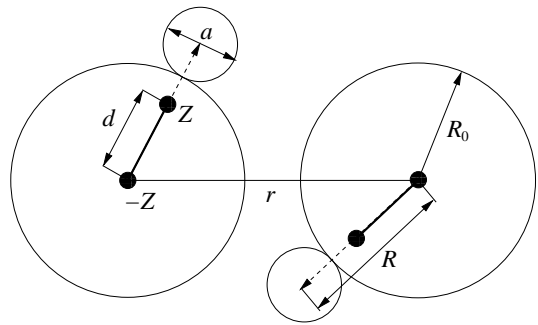} 
    \caption{Schematic representation of the toy model for two PE-colloid complexes at distance $r$ for 
      short chain lengths ($x \lesssim \pi$). One complex behaves like a dipole with charge $Z$ and length $d$ 
      (dumbbells). The short chain is modeled as a simple sphere sticking on the surface of the colloid in distance
      $R=R_0+a/2$ where $R_0$ is the colloid radius and $a$ is the effective diameter of the fluctuating chain.} 
    \label{sketch}
  \end{center}
\end{figure}

In order to obtain a better understanding of the small distance behavior
of the force for complexes with short chains, Fig.~\ref{force} (a), 
consider the toy model depicted in Fig.~\ref{sketch}.
Here the chains are modeled as hard spheres with effective diameter $a$.
Each sphere sticks to one colloid and can only move
on the surface of the macro sphere with radius $R_0$. One small sphere belonging to one
colloid can not penetrate into the other complex, and the center-to-center distance between
the colloids is $r$. In this case, the effective complex radius is $R=R_0+a/2$. 

As previously shown, complexes with short chains behave like rigid dipoles with length $d$ 
which is dependent of the chain contour length $l$.
Hence, in our toy model the complex has a dipole moment in direction from the center of the colloid
to the center of the small sphere with an effective length $d$
and charge $Z$. For distances $r>2(R_0+a)$ this model shows
a simple dipole-dipole attraction, whereas for smaller distances the dipole-dipole attraction 
is modified by the restricted number of available configurations.

\begin{figure} 
  \begin{center}
    \includegraphics[width=7cm, angle=-90]{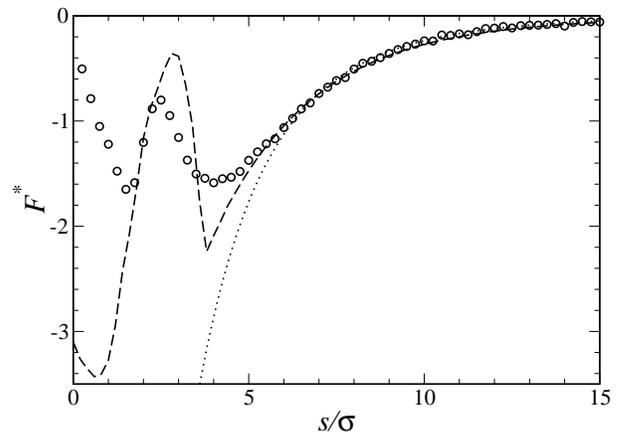} 
    \caption{Interaction force between two PE-colloid complexes with a
      short chain $x=0.79$ and $z=1$. Simulation
      results (circles) are compared to the numerical 
      solution of the toy model depicted in Fig.~\ref{sketch} (dashed line) and the analytical 
      result for the dipole-dipole interaction (dotted line).}
    \label{force_N10}
  \end{center}
\end{figure}

One can find the force between the complexes for this toy model by numerically solving the
partition sum and calculating the free energy as a function of the distance $r$. 
The results for a chain length $x=0.79$ is plotted in Fig. \ref{force_N10}
together with the corresponding simulation data. 
According to the results from the previous section we use
$\til{d}=6.3$. We use the diameter of the small spheres $a$ as a fit
parameter, which should in any case be of the order of the monomer size.
This parameter does not affect the dipole-dipole interaction, being only important
when the deviations from the pure dipole-dipole interaction sets in.
For $x=0.79$ and $z=1$ we used $\til a=3.6$. This model leads to a qualitative
agreement for the short range behavior of the force, as depicted in
Fig.~\ref{force_N10}: while the scale of the force is larger 
for the toy model, the minima and maxima are located at approximately the right distances. 
In other words, this means that the behavior observed for the force at small distances
and for small chains (a force with two minima) is at least partially explained by the
restriction of the number of available configurations of a system that is still essentially
composed by two interacting dipoles. 

It is also interesting to study the behavior of the
order parameter $\phi$ of our toy model solved for the length $x=0.79$
and $z=1$ also plotted in Fig. \ref{phi}. The qualitative behavior in
dependence of the distance is quite the same as in the simulation.
At $\til s\approx 4.5$ the dipole-dipole regime breaks
down and $\phi$ descends to negative values for closer distances $s$.
The negative value of $\phi$ means that in our toy model both small
spheres are located mainly between the macroions.
One concludes that due to the excluded volume interaction
of the small sphere/chain with the macrosphere the dipole-dipole
interaction is disturbed and the the spheres/chains  locate themselves
between the macroions, contrary to what happens in the dipole-dipole interaction
regime. 

\subsection{Long chains}

\begin{figure} 
  \begin{center}
    \includegraphics[width=6cm, angle=-90]{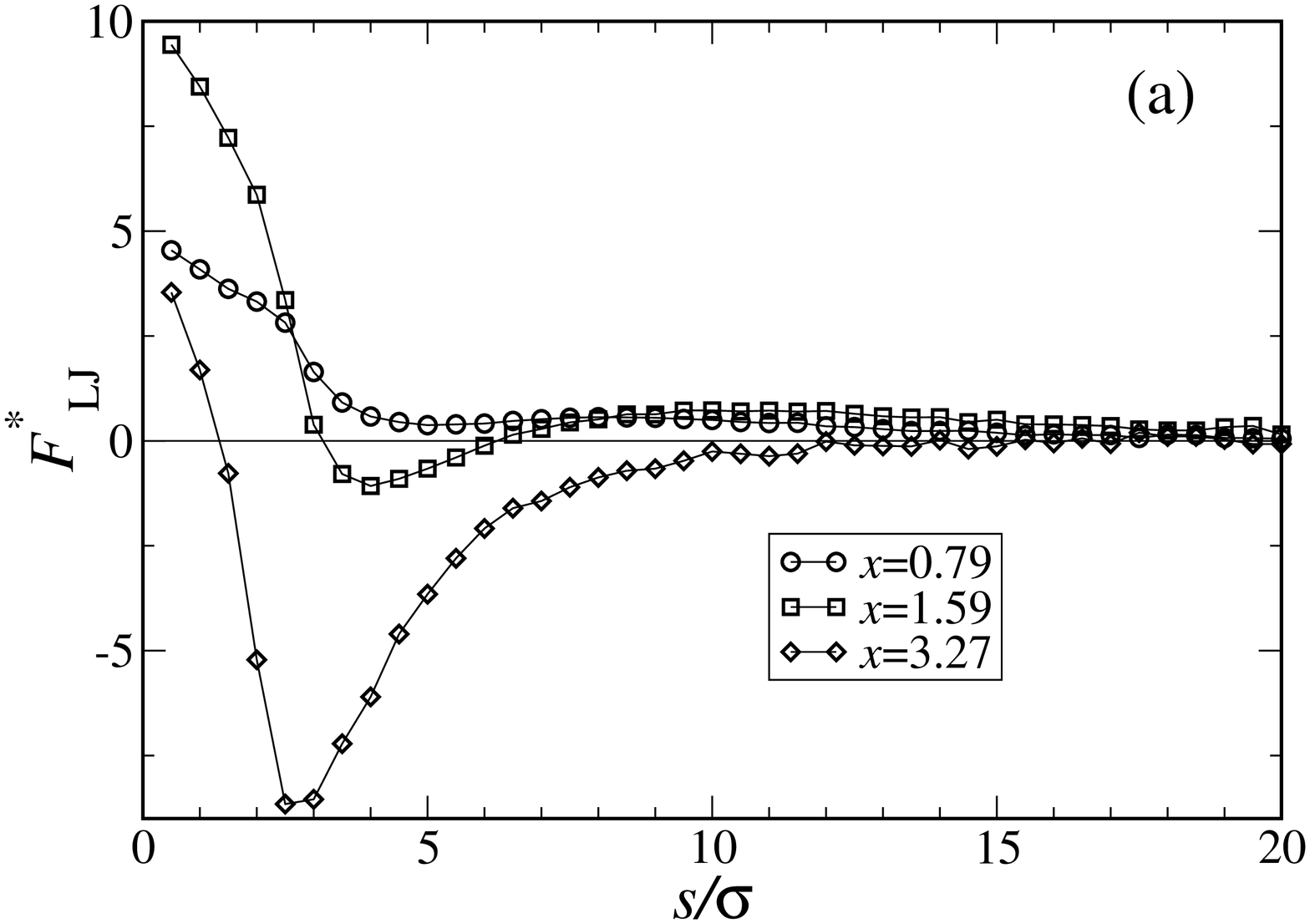} 
    \includegraphics[width=6cm, angle=-90]{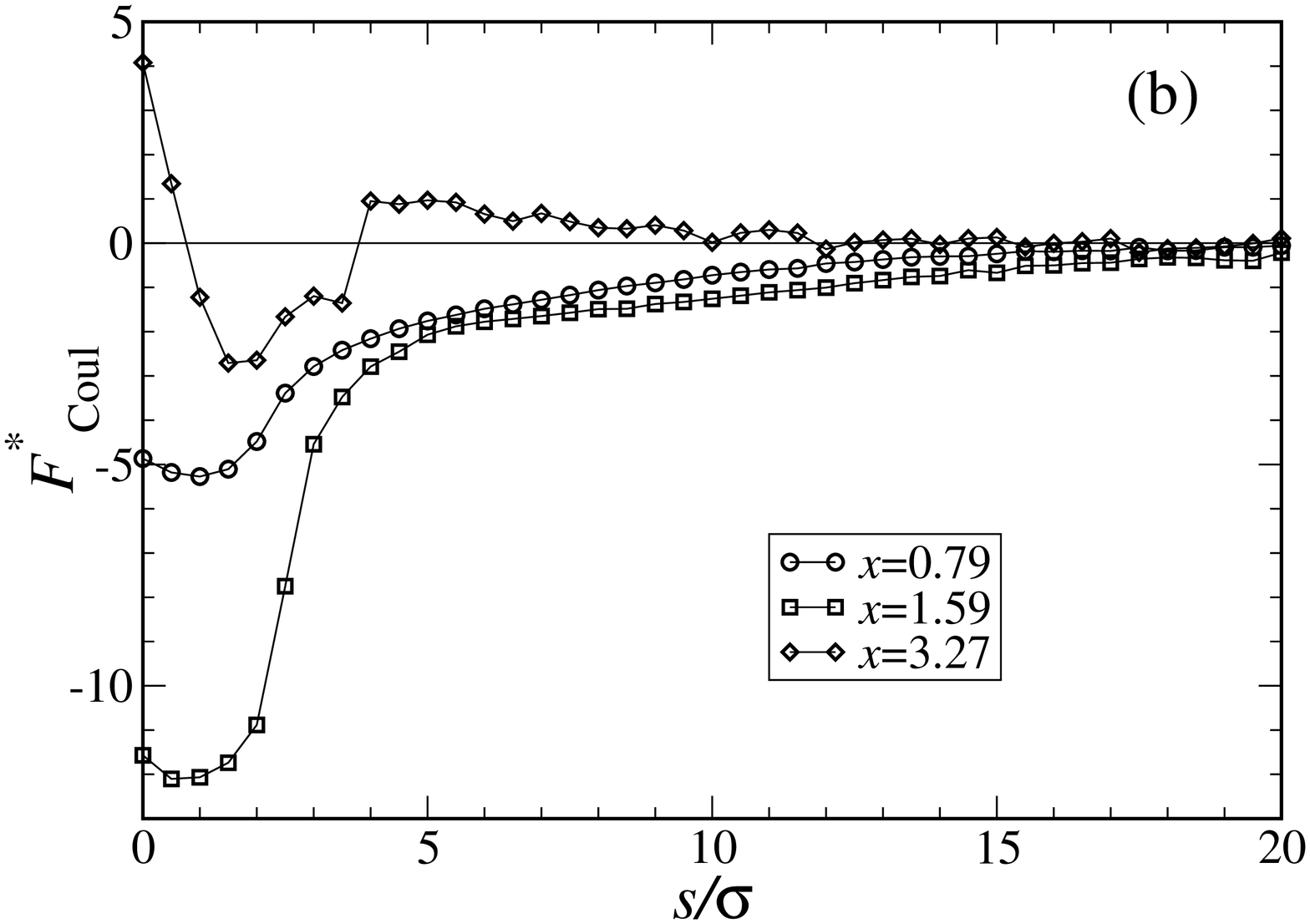} 
    \caption{Plot (a): simulation results of the entropic part of the 
      force $F^*_{\rm LJ}=\beta \, \sigma \, F_{\rm LJ}$ 
      defined in Eq.~(\ref{force.lj}) for different chain lengths and $z=1$. The colloid radius is $\til{R}_0=5.0$. 
      Plot (b): simulation results for the Coulombic part of the force 
      $F^*_{\rm C}=\beta \, \sigma \, F_{\rm C}$
      Eq.~(\ref{force.coul}) using the same parameters as in plot
      (a). Lines are guide to the eye.}
    \label{force_LJ}
  \end{center}
\end{figure}

\begin{figure} 
  \begin{center}
    \includegraphics[width=6cm, angle=-90]{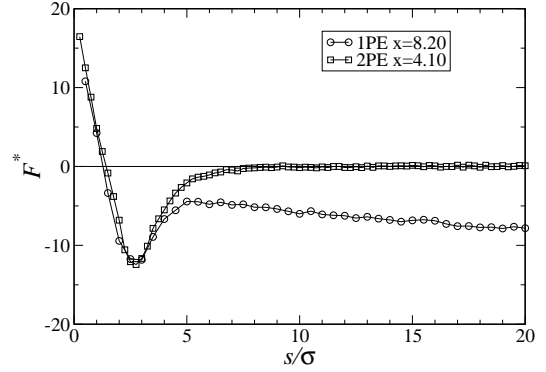} 
    \caption{Simulation results of the force between two colloids with $\til{R}_0=5.0$ carrying
      two chains of length $x=4.10$ (squares) and one chain with
      doubled length $x=8.20$ (circles). The monomer valence is
      $z=1$. Lines are guide to the eye. 
      }
    \label{force_onechain}
  \end{center}
\end{figure}

\begin{figure} 
  \begin{center}
    \includegraphics[width=6cm, angle=-90]{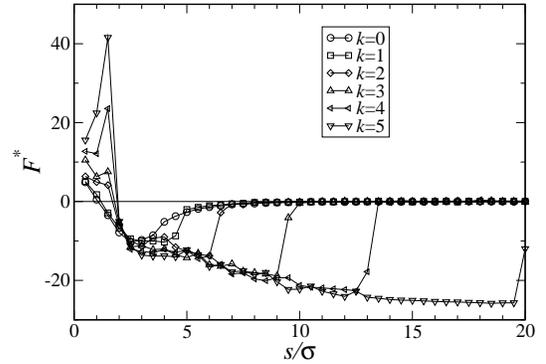} 
    \caption{Simulation results of the force between two PE-colloid complexes for a fixed chain
      length $x=3.27$, $z=1$, and different rigidity $k=\beta k_{\rm ang}$
      according to Eq.~(\ref{harm}). Lines are guide to the eye.}
    \label{force_stiff}
  \end{center}
\end{figure}

Increasing the chain length we observe a deep short ranged attraction as plotted
in Fig.~\ref{force} (b) which  is in agreement with previous
studies\cite{Akesson:jcp:89,Granfeldt:jpc:91,Podgornik:jcp:95}.
The depth of attraction initially grows with increasing chain
length while the range of attraction shrinks. Further increasing of
the chain length leads to a pure repulsion of the complexes.
The attraction is essentially of entropic origin, as the comparison between the
Coulomb force $F_{\rm C}$ and the entropic force $F_{\rm LJ}$ as
defined in Eq.~(\ref{force.coul}) and ~(\ref{force.lj}) shown in 
Fig.~\ref{force_LJ} demonstrates: for $x=3.27$ (large chain) the entropic force
clearly dominates over the Coulomb. For distances comparable to some
monomer lengths the chain is able to gain entropy by bridging the two colloids with some
parts of the chain, as seen for instance in snapshots Figs.~\ref{ss2}(e) and (f). 
In agreement with the absence of the dipole-dipole interaction between
the complexes, note that for chains with $x > \pi$, only zero or negative values 
of $\phi$ are observed: at short enough distances, the chains bridge between the colloids,
otherwise the complexes basically ignore each other and no long-range force is present.

Not surprisingly, for small distances the attraction of two
long chains is indistinguishable from the behavior of one chain with the combined length
of the two chains, as depicted in Fig.~\ref{force_onechain}, where the force 
between two complexes with chains with $x=4.10$ is compared with the force between to
colloids in the presence of only one chain with $x=8.20$.
The force is the same for small distances $\til{s} < 5$ where the entropic bridging regime
is found. The linear long ranged attraction of the single chain system is due to an energetic bridging
already observed and discussed in Refs.~\cite{Podgornik:jcp:95} and
\cite{Granfeldt:jpc:91}.

For complexes with very long chains only a 
pure repulsion is observable, see eg. $x=32.79$ in Fig. \ref{force}
(b). If the chain is long enough to
cover the whole surface of the macroion no bridging is possible due to
the electrostatic repulsion between the approaching chain and the
shell of monomers around the macroion.  A simple theoretical estimate
for the critical length when pure repulsion occurs is 
obtained by equating the available surface
of the spherical complex $A=4\pi R^{2}$ with the area $A=l\sigma$ of the
covering chain with length $l$ and thickness $\sigma$, which yields
\begin{eqnarray}
x_{\rm c}=2\pi R/\sigma.
\end{eqnarray}
We compared above estimate for two different complex radii
$R\approx 3.0$ and $R\approx 6.0$ with simulation results and found  
$x_{\rm c}\approx 15$ and $x_{\rm c}\approx 33$ which is very close to
the theoretical results $x_{\rm c}= 18.85$ and $x_{\rm c}= 37.70$.
See Fig. \ref{force} (b) for examples for the force between two
complexes with a colloid radius $\til R_{0}=5$.   

\subsection{Effects of chain stiffness or salt}
Increasing chain stiffness has an enormous influence on the
interaction for all chain lengths. The mechanism of interaction 
changes completely as now the rigid chains try to bridge between the
colloids to lower their bending energy which now is not negligible as
compared to the electrostatic energy. Simulation results for increasing
chains stiffness are shown in Fig.~\ref{force_stiff} 
for a chain length $x=3.27$. At a critical distance the chains bridge from one colloid 
to the other as shown in the snapshot in Fig.~\ref{ss2} (g). We checked that the isolated
complex with this length and bonding parameters shows always a complete adsorption of the
PE to the charged sphere and no one- or two-tail configurations. The functional behavior of the force
is linear for a wide range of the distance. This is similar to the interaction of two colloids 
with only one chain, see Fig.~\ref{force_onechain} discussed 
in Ref.~\cite{Podgornik:jcp:95} where the sharing of the chain was named
energetic bridging as one chain is then able to neutralize both
colloids and lower the electrostatic energy.
In our similar case with two chains the bridging is due to lowering of the
bending energy of the rigid chains.   
Adding salt shows no surprising behavior of the interactions for all
chain length. The shape and qualitative behavior of the force 
remains the same for low salt concentration, only the magnitude
of attraction decreases. As expected, the interaction of
short chains  is more affected by adding salt since the long-range
attraction is due to electrostatic interaction which are now
screened. Here differences already occur for a salt concentration of
$c_{s} \approx 0.05 {\rm M}$ and the long-ranged attraction vanishes
completely for 
$c_{s} \approx 0.2 {\rm M}$ (measured for $x=1.59$ and $z=1$).
The interaction of long chains has entropic origin and  remains
unaffected till $c_{s}$ exceeds values of approximately $ 0.1 {\rm M}
$ for a chain length $x=4.10$ and $z=1$. For values of $c_{s} \approx 1.0 {\rm
  M}$ still a weak attraction is observed.  
\section{Final remarks}
\label{conclusion}

In summary, we studied the polarizability and interaction of PE-colloid complexes for
the case where the chain is fully adsorbed on the macroion.
The polarizability is strong for small chains and small electrical
fields and has a maximum for chain length comparable to  
the circumference of the colloid. It can theoretically be described in terms of a thermally fluctuating 
dipoles with an effective length which depends on the chain length. For chains longer 
than the circumference of the macroion the polarizability is of the order of the polarizability 
of a classical conducting sphere with radius of complex size. The
interaction shows a van der Waals like attraction for short chains and
a deep short ranged attraction for midsized chains. Exceeding a
critical chain length $x_{c}$ the interaction is completely repulsive.
One could in principle try to explain the strong short-range attraction seen in the
simulations through the attractive interaction seen between
equally charged walls in the presence of polyelectrolytes\cite{bo1} or
counterions\cite{andre} (using the Derjaguin approximation to correct
for the spherical geometry). However, one should note that the Derjaguin
approximation is only valid in the limit where the colloidal particle
is much larger than the adsorbed chains, which is not the case in the
simulations presented here. 

One of the interesting conclusions that one can draw from this work concerns the stability
of colloidal suspensions which are stabilized by adsorption of polyelectrolytes. While for 
chains which are very long compared to the size of the particles
$x>x_{c}$ this is indeed an effective 
mechanism of stabilization (as shown in Fig.~\ref{force} for $x=32.79$, 
where the forces are always repulsive or zero), complexes with chains with a length
comparable with the colloidal size exhibit a short range attraction due to the so-called bridging
between the particles. However, for chains that are smaller than the colloid, a 
long-range attraction is present and one should expect such suspensions to behave like
dipolar fluids\cite{ze-maria}. This shows that the stabilization of colloidal suspensions by 
polyelectrolyte adsorption is strongly dependent on the chain size relative to the colloidal
size: for long chains the suspensions are always stable (only repulsive forces between the
particles), while for mid-sized and short chains there is attraction between the complexes and
a salting-out can occur. Although our studies were essentially done for isolated, 
strongly charged and symmetric complexes these conclusions should hold in more
realistic situations where the chains have some polydispersity and
the complexes are at relatively high concentrations, since the only key point
is the existence of the complexes.

\acknowledgments 
The authors thank H.\ L{\"o}wen and C.\ Likos for a critical reading of the manuscript.
JD \ thanks  O.\ Farago and M.\ Sabouri-Ghomi for useful discussions.  
This research was in part supported by the National 
Science Foundation under Grant No. PHY99-72246 and by the MRSEC
program of the NSF under award  no. DMR00-80034. JD acknowledges
financial grant from the DAAD Doktorandenstipendium and support from
the DFG under "Schwerpunkt Polyelektrolyte".

\end{document}